\documentclass[11pt]{article}
\usepackage{amssymb,cite}
\usepackage{epsfig}

  \let\n=\nu

\let\C=\Chi

\def\nn{\nonumber} \def\bd{\begin{document}} \def\ed{\end{document}}
\def\ds{\documentstyle} \let\fr=\frac \let\bl=\bigl \let\br=\bigr
\let\Br=\Bigr \let\Bl=\Bigl 
\let\bm=\bibitem
\let\na=\nabla
\let\pa=\partial \let\ov=\overline 
\newcommand{\be}{\begin{equation}} 
\newcommand{\ee}{\end{equation}} 
\def\ba{\begin{array}}
\def\ea{\end{array}}
\def\ft#1#2{{\textstyle{{\scriptstyle #1}\over {\scriptstyle #2}}}}
\def\fft#1#2{{#1 \over #2}}
\def\del{\partial}
\def\vp{\varphi}
\def\st#1{{\scriptstyle #1}}
\def\sst#1{{\scriptscriptstyle #1}}

\def\oneone{\rlap 1\mkern4mu{\rm l}}
\def\td{\tilde}
\def\wtd{\widetilde}
\def\ie{{\it i.e.\ }}
\def\dalemb#1#2{{\vbox{\hrule height .#2pt
        \hbox{\vrule width.#2pt height#1pt \kern#1pt
                \vrule width.#2pt}
        \hrule height.#2pt}}}
\def\square{\mathord{\dalemb{6.8}{7}\hbox{\hskip1pt}}}

\def\cramp{\medmuskip = 2mu plus 1mu minus 2mu}
\def\cramper{\medmuskip = 2mu plus 1mu minus 2mu}
\def\crampest{\medmuskip = 1mu plus 1mu minus 1mu}
\def\uncramp{\medmuskip = 4mu plus 2mu minus 4mu} 

\font\smallfont = cmr9

\newcommand{\ho}[1]{$\, ^{#1}$}
\newcommand{\hoch}[1]{$\, ^{#1}$}
\newcommand{\bea}{\begin{eqnarray}} 
\newcommand{\eea}{\end{eqnarray}} 
\newcommand{\ra}{\rightarrow}
\newcommand{\lra}{\longrightarrow}
\newcommand{\Lra}{\Leftrightarrow}
\newcommand{\ap}{\alpha^\prime}
\newcommand{\bp}{\tilde \beta^\prime}
\newcommand{\tr}{{\rm tr} }
\newcommand{\Tr}{{\rm Tr} } 
\def\0{{\sst{(0)}}}
\def\1{{\sst{(1)}}}
\def\2{{\sst{(2)}}}
\def\3{{\sst{(3)}}}
\def\4{{\sst{(4)}}}
\def\5{{\sst{(5)}}}
\def\6{{\sst{(6)}}}
\def\7{{\sst{(7)}}}
\def\8{{\sst{(8)}}}
\def\n{{\sst{(n)}}}
\def\cA{{{\cal A}}}
\def\cF{{{\cal F}}}
\def\tV{\widetilde V}
\def\tW{\widetilde W}
\def\tH{\widetilde H}
\def\tE{\widetilde E}
\def\tF{\widetilde F}
\def\tA{\widetilde A}
\def\im{{{\rm i}}}
\def\jm{{{\rm j}}}
\def\km{{{\rm k}}}

\def\tY{{{\wtd Y}}}
\def\ep{{\epsilon}}
\def\vep{{\varepsilon}}
\def\R{\rlap{\rm I}\mkern3mu{\rm R}}
\def\bD{{{\bar D}}}
\def\R{{{\mathbb R}}}
\def\C{{{\mathbb C}}}
\def\H{{{\mathbb H}}}
\def\CP{{{\mathbb C}{\mathbb P}}}
\def\RP{{{\mathbb R}{\mathbb P}}}
\def\Z{{{\mathbb Z}}}
\def\bA{{{\mathbb A}}}
\def\bB{{{\mathbb B}}}
\def\bx{{\bf x}}
\def\wtd{\widetilde}
\newcommand{\NP}{Nucl. Phys. }
\newcommand{\tamphys}{\it George P. \& Cynthia W. Mitchell
Institute for Fundamental Physics,\\
Texas A\&M University, College Station, TX 77843, USA}

\newcommand{\damtp}{\it DAMTP, Centre for Mathematical Sciences,\\
 Cambridge University, Wilberforce Road, Cambridge CB3 OWA, UK}

\newcommand{\auth}{W. Chen,\hoch{\ddagger} Z.-W. Chong\hoch{\ddagger} 
G.W. Gibbons\hoch{\sharp}, 
H. L\"u\hoch{\ddagger} and   C.N. Pope\hoch{\ddagger 1}}

\begin{document}
\begin{flushright}
\hfill{MIFP-05-04\ \ \ \ DAMTP-2005-12\ \ \ \
{\bf hep-th/0502077}}\\
{February 2005}
\end{flushright}


\begin{center}
{ \large {\bf Ho\v rava-Witten Stability: eppur si muove}}

\vspace{10pt}
\auth

\vspace{10pt}
{\hoch{\ddagger}\tamphys}

\vspace{10pt}
{\hoch{\sharp}\damtp}

\vspace{20pt}

\underline{ABSTRACT}
\end{center}

     We construct exact time-dependent solutions of the supergravity
equations of motion in which two initially non-singular branes, one
with positive and the other with negative tension, move together and
annihilate each other in an all-enveloping spacetime
singularity. Among our solutions are the Ho\v rava-Witten solution of
heterotic M-theory and a Randall-Sundrum I type solution, both of
which are supersymmetric, \ie BPS, in the time-independent case.  In
the absence of branes our solutions are of Kasner type, and the source
of instability may ascribed to a failure to stabilise some of the
modulus fields of the compactification.  It also raises questions
about the viability of models based on some sorts of negative tension
brane.

{\vfill\leftline{}\vfill
\vskip 5pt
\footnoterule
{\footnotesize \hoch{1} Research supported in part by DOE grant
DE-FG02-95ER40899}}

\pagebreak
\setcounter{page}{1}

\vfill\eject

\section{Introduction}

One of the most popular current models relating M-theory to
phenomenology is that of Ho\v rava and Witten \cite{H-W}, in which an
eleven-dimensional spacetime is the product of a compact Calabi-Yau
manifold with a 5-dimensional spacetime consisting of two parallel
3-branes or domain walls, one with negative tension and one with
positive tension.  Alternatively one may think of the five-dimensional
reduced spacetime as the warped product of Minkowski spacetime with an
interval, \ie ${\mathbb E}^{3,1} \times S^1/{\mathbb Z}_2$.

    If one performs a generalised dimensional reduction of the
eleven-dimensional theory, of the kind introduced in \cite{lalupo} in
which the 4-form has non-vanishing flux on a Calabi-Yau internal
space, one obtains a five-dimensional supergravity theory
\cite{Ovrutetal} which admits an exact static supersymmetric solution
of the form
\bea
ds_5^2 &=& \wtd H \, (-dt^2 + d \bx ^2)  + \wtd H^4\,
d\td y^2\,,\nn\\
\wtd H &=& 1 + \td k\, |\td y|\,,
\qquad \phi = -3\log \wtd H\,,\label{solution1}
\eea
where $\td k$ is a constant.  The scalar field $\phi$ characterises
the size of the internal Calabi-Yau space.  The equations of motion
for the metric and $\phi$ may be consistently obtained from the
Lagrangian
\be
{\cal L}_5 = \sqrt{-g}(R - \ft12 (\del\phi)^2 - m^2\, e^{2\phi})
\,,\label{lagrangian}
\ee
with $\td k^2= 2m^2/3$, where the exponential potential is the remnant
of the 4-form field strength.  Note that if the solution is lifted
back to $D=11$, it can be viewed, in the orbifold limit, as an
intersection of three equal-charge M5-branes \cite{lalupo}.

   In the Ho\v rava-Witten picture a second domain wall is introduce,
at $y=L$, by taking $y$ to be periodic with period $2L$, such that
$y=L$ is identified with $y=-L$.  Furthermore, one makes the ${\mathbb
Z}_2$ identification $y \leftrightarrow -y$.  This solution has been
proposed as a model for our universe \cite{Ovrutetal}, and so the
question of its stability is of obvious interest.

   The Ho\v rava-Witten model based on the solution (\ref{solution1})
is not completely physically realistic because it is static, while our
expanding universe is time dependent. For that reason, many attempts
have been made to incorporate the Ho\v rava-Witten model into a
cosmological, so-called brane-world, scenario, in which the positions
of the 3-branes or domains walls are not fixed but allowed to move in
time. Notable among these attempts are the Ekpyrotic scenario
\cite{Ek}, in which the big bang is ascribed to the collision of an
external brane with our universe, and the Cyclic model \cite{cyclic},
in which the distance between the two branes oscillates in time.

Brane scenario cosmologies, particularly those based on collisions,
represent a rather radical departure from previous models, and offer a
novel perspective on many long-standing problems and puzzles, but they
also present new ones of their own.  Understandably there has been a
great deal of interest in them.  However, much of the recent work has
been based on effective four-dimensional theories, typically involving
a so-called radion field.  Once the theory has been brought to this
form it is almost indistinguishable from conventional four-dimensional
models involving scalar fields; only the names have been changed.  If
brane scenarios are to be observationally tested it must be via
features that are essentially higher dimensional.  In particular, in
the case of collision models we need a much better understanding of
the higher-dimensional collision dynamics derived from a consistent
underlying framework such as M-theory. Since at present we lack a
complete theoretical formulation of M-theory, any such further
understanding at present must come from incomplete or approximate
theories such as low-energy supergravity limits or Dirac-Born-Infeld
actions. This paper is concerned with the former approach. For some
ideas on topology and signature change using the latter, see
\cite{Ishibashi}.

In the light of the comments made above, it is clearly worthwhile to
obtain time-dependent solutions of the equations of motion coming from
(\ref{lagrangian}), and to relate the analysis of stability to the
suggestions of \cite{Ek} and \cite{cyclic}. In a recent paper
\cite{glp}, exact solutions of the supergravity equations of motion
representing colliding D3-branes of Type IIB theory moving in ten
spacetime dimensions were obtained.\footnote{Time-dependent
four-dimensional charged black-hole solutions in a de Sitter
background were earlier obtained in \cite{kastra}, and generalistions
to higher-dimensional charged black holes with a pure exponential
scalar potential were found in \cite{makshi}.}  By dimensionally
reducing these solutions, a class of five-dimensional time-dependent
solutions was obtained for a similar Lagrangian to (\ref{lagrangian})
(but with a different power of the dilaton exponential potential)
\cite{glp}. The static solutions were supersymmetric, but the
time-dependent generalisations represent a positive and a negative
tension brane moving towards one another, and leading to the complete
disappearance of the universe in a spacetime singularity \cite{glp}.
In this paper, we shall construct exact solutions of the equations of
motion coming from precisely the Lagrangian (\ref{lagrangian}) of the
heterotic brane model of \cite{Ovrutetal}, with the same type of
time-dependent properties. Their existence clearly raises questions
about the validity of the general belief that supersymmetric ground
states should be stable.

   We should say at the outset that in what follows we shall be
concerned only with the exact classical supergravity equations of
\cite{Ovrutetal}, and any effects arising from quantum considerations,
which can induce potentials for the various massless scalar fields in
the theory (see, for example, \cite{KKLT} and \cite{becukr,Beckers}), are not
taken into account. Clearly, such potentials would modify considerably
the discussion that follows.

   We shall begin by obtaining a five-dimensional time-dependent
domain-wall solution for precisely the Lagrangian (\ref{lagrangian})
that arose in the heterotic brane model of \cite{Ovrutetal}.  We shall
include a detailed discussion of the nature of the time dependence,
and also we shall make analogy with the issue of the stability of
ordinary Minkowski spacetime.  In subsequent sections we shall obtain
more general classes of time-dependent domain-wall solutions in
arbitrary dimensions, and we shall also discuss the brane-source terms
that arise in all the cases.

\section{Time-Dependent Heterotic Brane in $D=5$}

\subsection{The local time-dependent solution}

    As well as admitting the static 3-brane solution
(\ref{solution1}), we find that the five-dimensional theory described
by the Lagrangian (\ref{lagrangian}) also admits a time-dependent
3-brane solution, given by
\bea
ds_5^2 &=& H^{1/2} \, (-dt^2 + d\bx^2)  + H\, dy^2\,,\nn\\
H &=& h\, t + k\, |y|\,,\qquad \phi = -\ft32 \log H\,,\label{tsol1}
\eea
where $k^2 = 8 m^2/3$, and $h$ is an arbitrary constant.

   Note that if we turn off the time dependence (by setting $h=0$),
the relation to the previous static solution is seen by making a
coordinate transformation of the form $y= \td y^2$.  Substituting this
into (\ref{tsol1}), we recover (\ref{solution1}), after some simple
constant rescalings.  If, on the other hand, we set the parameter $m$
in the Lagrangian to zero, the solution describes a Kasner universe.

     When we lift the solution back to $D=11$, the metric becomes
\be
ds_{11}^2 = H^{-1/2} (-dt^2 + d\bx^2) + 
H^{1/2}\, ds_{CY_6}^2 + dy^2 \,,\label{d11sol}
\ee
The static solution, in the orbifold limit, can be viewed as an
intersection of three equal-charge M5-branes \cite{lalupo}.  Turning
off the brane charge, the time-dependent metric describes a direct
product of a ten-dimensional Kasner universe and a line segment.

\subsection{Local static Killing vectors and Killing horizons}
\label{timesec}

   Although we have presented the solution (\ref{tsol1}) in an
ostensibly time-dependent form, one might worry that in fact it was
not truly time-dependent at all, since locally one can eliminate the
time dependence by means of a coordinate transformation.  If we
temporarily drop the modulus sign around $y$ in (\ref{tsol1}), then
the coordinate transformation from $t$ and $y$ to $\td t$ and $r$
given by
\be
dt=d\td t - \fft{h r^{1/2}}{k^2\, f}\, dr\,,\qquad
H=r\,,\qquad f= 1-\fft{h^2\, r^{1/2}}{k^2}
\ee
transforms the solution into the static form 
\bea
ds_5^2 &=& r^{1/2}\, \Big( -f\, d\td t^2 + d\bx^2 + r^{1/2}\,
\fft{dr^2}{k^2\, f}\Big)\,,\nn\\
\phi &=& -\ft32 \log r\,.\label{localstatic}
\eea
This can be recognised as a black 3-brane, with an horizon at $f=0$.
However, as we shall discuss below, the introduction of the modulus
sign on $y$ in (\ref{tsol1}) changes the conclusion 
completely.

   First, it is useful to continue temporarily with the modulus sign
omitted, and to look at the Killing symmetries of the solution.
It depends on $t$ and
$y$ only though the combination $H=ht+ky$, and so if one adopts $H$ and 
another combination of $t$ and $y$, say $v=ht-ky$, as new coordinates,
then the solution is independent of $v$. Thus 
\be
K= K^\mu { \del \over \partial x ^\mu }=
 k\, \fft{\del}{\del t}  -h\, \fft{\del}{\del y} \,, 
\ee
is a Killing vector field which lies in the hypersufaces $H={\rm
constant}$, and which has norm
\be g(K,K)\equiv g_{\mu \nu} K^\mu K^\nu = -k^2 H^{1 \over 2} + h^2H\,.
\ee
In other words
\be
K(H)\equiv  K^\mu \partial _\mu H=0\,.
\ee
Thus, if one moves such that $ht+ k y= {\rm constant}$, the domain
wall appears to be independent of the ``time'' coordinate $v$.  If $H
< {k^4 \over h^4}$, then the Killing vector is timelike and $v$ is a
genuinely timelike coordinate, but if $H > {k^4 \over h^4}$ it is
spacelike, and $v$ becomes a spacelike coordinate. On the hypersurface
$N$ given by $H= {k^4 \over h^4}$, the Killing vector field is null,
\be
g(K,K)= g_{\mu \nu} K^\mu K^\nu =0\,.
\ee  
Thus the hypersurface $N$ is itself null, and the orbits of
$K$ were its null generators. Such hypersurfaces 
are called Killing horizons by Carter \cite{Carter}.

   As mentioned above, our domain-wall solution (\ref{tsol1}) is in
fact genuinely time dependent, despite being locally static, and the
time dependence is not a mere coordinate artefact.  The crucial point
is the inclusion of the modulus sign in (\ref{tsol1}), which
introduces a physical domain wall at $y=0$. If we sit on this wall,
the metric is explicitly time dependent because the function $H$ is
not constant at the wall.  In other words the domain wall does not
move along the orbits of the Killing vector $K$.

    In fact the hypersurfaces $H={\rm constant}$ have a kink at $y=0$.
The vector $K$ suffers a jump in slope at $y=0$, and thus the
one-parameter group of translations which it generates is
discontinuous at $y=0$.

   The fact that the metric is locally static outside a domain wall is
a rather general phenomenon.  For example in the thin-wall
approximation, a four-dimensional domain wall spacetime consists of
two pieces of flat Minkowski spacetime inside the timelike hyperboloid
\be
x^2+y^2+z^2-t^2 = a^2\,,
\ee 
which are glued back to back, thus compactifying spacetime.  On either
side of the domain wall the metric is static, but the domain wall
itself is moving with respect to any of the doubly-infinite number of
static coordinate systems related by $SO(3,1)$ transformations on
either side. However there is no global static coordinate system
\cite{Gibbons}. The domain wall itself, and the universe containing
it, are expanding (or contracting), and in terms of a proper-time
parameter $\tau$ the expansion is exponential. In other words, the
universe is inflating because of the presence of the domain wall.  In
fact the induced metric on the domain wall is precisely that of 2+1
dimensional de Sitter spacetime,
\be
ds^2=-d\tau^2 + a^2\, \cosh^2 {\tau \over a} \,
( d \theta ^2 + \sin^2\theta d \phi ^2 )\,.
\ee
Analogously, our 3-brane is also a thin domain wall solution, which is
locally static outside the wall.  If we could smooth out the kink and
replace the $|y|$ profile in (\ref{tsol1}) by a smooth trough, then
the solution would no longer be even locally static.

   Having established that our 3-brane solution is indeed genuinely
time-dependent, it is appropriate to discuss the nature of its time
evolution in more detail.  This discussion is closely analogous to the
evolution of the time-dependent 3-brane considered in \cite{glp}.

\subsection{Global structure of the time-dependent
Ho\v rava-Witten spacetime}

   In this section we shall discuss the global structure of the
time-dependent generalisation of the Ho\v rava-Witten model.  This
corresponds to taking the time-dependent solution (\ref{tsol1}) and
passing to the case of the $S^1/{\mathbb Z}_2$ orbifold.  Thus we
consider a solution of the form (\ref{tsol1}) for $-L<y<L$. The
solution is then extended outside this interval by demanding that it
be periodic with period $2L$. This leads to a negative-tension brane
located at $y=0$ and a positive-tension brane located at $y=L$. The
interval $0\le y \le L$ may be though of as $S^1/{\mathbb Z}_2$ ,
where the $S^1$ occupies $-L\le y \le L$ and the ${\mathbb Z}_2$
action is $y \rightarrow -y$.

   Now if, as in \cite{glp}, $h$ is taken to be negative, $H$ will be
positive but decreasing, for all negative values of $t$ and the
spacetime occupies the strip $0\le y \le L$ when $t<0$.  Evidently the
proper length of the interval is time dependent, and the universe is
contracting, but not exponentially.  The proper length of the interval
is
\be
\ell =\int _0^L dy H^{1/2} = \fft{2}{3k}\, \Big[ (ht + k L)^{3/2} - 
(ht)^{3/2} \Big]\,.
\ee
At large negative $t$ we have
\be
\ell \approx (ht)^{1/2}L\,, 
\ee
and at $t=0$, the proper length is $\ell={2 \over 3}  k^{1/2} L^{3/2}$.

   For $t$ positive, $H$ vanishes along the straight line
\be
-ht=ky\,.
\ee
This represents a singularity which starts from the negative tension
brane and moves towards the positive tension brane, reaching it at
time $t={kL \over (-h)} $. The spacetime cannot be extended beyond
this, because the scalar field $\phi$ diverges, giving rise to a
curvature singularity.  This can be seen from the calculation
presented in (\ref{riemann}) in the appendix; the curvature clearly
diverges at $H=0$.  It can also be seen from the metric (\ref{d11sol})
in $D=11$, which becomes complex when $H$ is less than zero.

   To summarise, the separation of the two 3-branes decreases
monotonically as $t$ increases towards zero, but before they actually
collide a power-law curvature singularity develops on negative-tension
brane at $t=0$, which spreads out and eventually envelopes the entire
spacetime including the positive-tension brane, at time $t={kL \over
(-h)} $.

   The norm of the local Killing vector is
\be
H^{1/2} ( h^2 ( ht - ky )^{ 1/2} -k^2)\,.
\ee
For large negative $t$ the Killing field is spacelike.  For small
negative $t$ is becomes timelike near the negative tension brane. For
$L<{ k^3 \over h^2}$, it has become timelike along the entire interval
before $t=0$.

   The  Einstein conformal frame metric induced on the  branes  is
\be   
ds^2 = H (-dt^2 + d \bx^2 )\,.
\ee
For the brane at $y=0$ and  negative values of $t$, the proper time on
the brane is  $\tau = {2 \over 3} (ht) ^{ 1/2 } t$. The metric becomes  
\be
ds^2 =-d\tau^2 + a^2(\tau) d \bx ^2 \,, \qquad a(\tau) \propto 
\tau^{ 1/3}\,.
\ee  
The contraction towards the Big Crunch at $\tau=0$ is power law, with
the expected power due to a massless scalar field.  A similar
contraction takes place on the positive tension brane, simply shifted
in time by an amount ${ kL \over (-h)}$.

   The natural interpretation of the solution is that the collision of
the negative tension brane with the positive tension brane brings
about the complete annihilation of the universe.

    From the $D=11$ point of view, the interpretation is different in
detail but the same in essence.  As we can see from (\ref{d11sol}),
the coordinate $y$ itself measures proper distance in $D=11$.  The
solution can be viewed as an M5-brane wrapped on the supersymmetric
2-cycles of $CY_6$ in a Kasner spacetime.  If we turn off the brane
charge $k$, the metric is a direct product of a ten-dimensional Kasner
universe and a line segment.  Thus the coordinate $y$ describes
precisely the original Ho\v rava-Witten line-segment of $S_1/Z_2$,
except that in the original Ho\v rava-Witten picture, the
ten-dimensional spacetime is (Minkowski)$_4\times CY_6$ instead of a
Kasner universe with the spatial sections being ${\mathbb E}^3\times
CY_6$.

     When the M5-brane charge is turned on, the wrapped M5-branes are
perpendicular to the $y$ direction. The distance between the branes
stays fixed in $D=11$, since $y$ is a proper-distance coordinate.  The
metric becomes singular when $H=0$, in which case the volume of the
Calabi-Yau manifold shrinks to zero.  The metric cannot be extended to
$H<0$ since it would then become complex.  If there is only a single
brane, so that $y$ is an infinite line, then there are always regions
of $y$, for any given $t$, for which the metric is well-defined.
However in the Ho\v rava-Witten model, where $y$ is a line segment
with a brane at each end, the universe is totally annihilated at the
time $t=k L/(-h)$, after which $H$ becomes negative for all $y$ in the
interval $S^1/{\mathbb Z}_2$.

    We have shown that the usual static heterotic five-dimensional
3-brane solution found in \cite{Ovrutetal} has a time-dependent
generalisation that is highly singular.  It describes the subsequent
time evolution if one sets the two 3-branes in motion towards each
other with a small velocity.  It is appropriate therefore to address
in more detail the question of whether this represents an inherent
instability of the static 3-brane.  In fact a useful analogy is to
consider first certain singular time-dependent generalisations of the
Minkowski metric itself, and the question of whether their existence
is indicative of an instability of flat spacetime.  We address this
question in the following subsection, as a prelude to discussing the
stability of the Ho\v rava-Witten spacetime itself in the subsequent
subsection.

\subsection{Stability versus instability of flat spacetime}

      It may be useful to begin by recalling some elementary and
widely appreciated facts about the stability of flat spacetime.  Of
course there is no doubt that in theories with fields which can carry
only positive energy, flat spacetime is stable against perturbations
of finite total energy, such as might be produced in a terrestrial
laboratory. Cosmologically however, the situation is different because
there is no obvious reason why we should impose a condition of finite
total energy, and indeed to do so would seem to violate the so-called
Cosmological Principle, which rules out privileged spatial locations
in the universe.  A perturbation would need to fall off quite sharply
away from where it is largest in order to be of finite total energy.
In fact if the flat spacetime were spatially compact, for example if
the spatial sections were tori, then presumably all perturbations are
of finite (but possibly vanishing) total energy. But this raises a
problem with stability.

    Consider, for example, the exact Kasner solutions of the vacuum
Einstein equations,
\be
ds^2 = -dt^2 + t^{2p_1} dx^2 + t^{2p_2} dy^2 + t^{2 p_3} dz^2\,, 
\ee
where $p_1,p_2,p_3$ are constants such that
\be
p_1+p_2+p_3=1=p_1^2 +p_2^2 + p_3^2\,.
\ee
Unless one of the $p_i$ is equal to 1, these metrics have a
singularity at $t=0$.  If we set $t=1 -t^\prime$, then the metric near
$t=1$ starts out looking like a small deformation of the flat metric,
with a small homogeneous mode growing linearly with $t^\prime$.
Ultimately, however, non-linear effects take over and the universe
ends in a Big Crunch at $t^\prime =1$, \ie $t=0$.

  This instability is universal in gravity theories, and is closely
related to the modulus problem in theories with extra dimensions.
Consider, for example, the exact ten-dimensional Ricci-flat metric
\be
ds^2 = t^ { { 1/ 2}} (-dt ^2 + d {\bf x} ^2 ) + 
t^{1/ 2}  g_{mn}(y) dy^mdy^n,
\ee
where $g_{mn}(y)$ is a six-dimensional metric on a Calabi-Yau space
$K$.  This starts off at $t=1$ looking like ${\mathbb E} ^{3,1} \times
K$ with a small perturbation growing linearly in $t^\prime= 1-t$.
However by the time it reaches $t^\prime =1$, $t=0$, the solution has
evolved to give a spacetime singularity. From the point of view of the
four-dimensional reduced theory, the logarithm of the volume of the
Calabi-Yau behaves like a massless scalar field -- the modulus field
which is sometimes thought of as a kind of Goldstone mode for a
spontaneously-broken global scaling symmetry. This causes an isotropic
expansion or contraction of the three spatial dimensions, with the
scale factor $a(\tau)$ going like $\tau^{1 \over 3}$, which is what
one expects for a fluid whose energy density equals its pressure.

   It is clear that no argument based on the local energy density of
the effective three-dimensional theory, or on the fact that ${\mathbb
E} ^{3,1} \times K$ is supersymmetric, \ie admits Killing spinors, can
eliminate this source of instability.  It is intrinsic to the
situation being considered, and to the fact that nothing sets the
scale of the Calabi-Yau manifold.  The same is true of the Kasner type
instability of flat space described previously. If one imagines
reducing the Kasner solution on all the spatial dimensions, leaving
just the time direction, one has three modulus fields or Goldstone
modes, corresponding to the three arbitrary length scales.  Note
further that if we do identify the spatial directions, any global
energy or supercharge functional will automatically vanish, because
they may be expressed as boundary terms of a space with no boundary.
Thus arguments based on Witten identities, {\it etc}., cannot be
applied in this case.

\subsection{Stability of Ho\v rava-Witten spacetime}

With the discussion above in mind, we can return to the issue of
stability for the brane solutions. The situation is rather similar to
the Kasner instability of flat spacetime. Indeed our solution reduces
to a Kasner type solution if we set $k=0$, \ie in the absence of the
branes.

     As always in physics, whether we say a system is stable or
unstable depends upon what boundary conditions we are prepared to
allow in the past.  
In the case of an environmental science like
cosmology this is well understood to be problematic. It becomes even
more so in brane cosmology since imposing a boundary condition in the
past is tantamount to declaring what influences are allowed on our
universe from ``outer space,'' that is in, our case, from higher
dimensions. What our calculations clearly show is that if no such
restrictions are imposed, then the Ho\v rava-Witten spacetime is
certainly unstable.  Any argument showing it to be stable must
therefore be an argument based on a theory about the initial
conditions.  In this respect, the parallel with the debate between
Catastrophists and Uniformitarians among nineteenth century geologists
trying to unravel the history of the earth is rather striking
\cite{Whewell}.

\subsection{Relation to Chamblin-Reall's work}

In this section we shall review the catastrophe 
awaiting the denizens of a Ho\v rava-Witten brane-world
in the manner of Chamblin and Reall \cite{ChamblinReall}. They focus
on the locally static form of the metric and regard  the 3-branes
as moving in it, rather than using the co-moving description we have
adopted. One starts from the full Kruskal-type extension
of the locally static manifold. This is constructed by first introducing
advanced and retarded Eddington-Finkelstein coordinates $(u,v)$ for
the locally-static metric given in (\ref{localstatic}):       
\be
dv= k\, d\td t + {r^{1/4}\, dr \over f}\,,
   \qquad dv= k\, d\td t - { r^{1/4}\, dr \over f}\,.
\ee 
Next, Kruskal coordinates are defined by 
\be
V= e^{-\kappa v}\,,\qquad U=  e^{\kappa u}\,.
\ee 
where $\kappa$ is a constant to be determined shortly.

    The metric now takes the form
\be
ds ^2 = { r^{1\over 2} f \over k^2 \kappa ^2 UV } \,
dU dV + r^{ 1\over 2} d {\bf x} ^2 \,,
\ee 
where $r$ should be regarded as a function of $UV$. The 
$SO(1,1))_0 \equiv {\mathbb R}_+ $ symmetry\footnote
{ The subscript $_0$ denotes the 
component connected to the identity. The group ${\mathbb R}_+$ is the group
of positive reals under multiplication.}  of the metric is now manifest 
since it is invariant under the boosts
$U \rightarrow \lambda U$, $V \rightarrow  \lambda ^{-1}  V$,
 whose orbits in the $U-V$ plane
are hyperbolae $UV={\rm constant} \Leftrightarrow r 
\equiv H = {\rm constant}$.
The thus-far arbitrary constant $\kappa$ is now chosen so that 
$f \over UV$ is non-zero and  analytic in $UV$ 
on the past and  future horizons, which are at $UV=0 \Leftrightarrow
r= {k^ 4 \over h^4}$. This implies that we should take 
\be
\kappa = {h^4 \over 4 k^4}\,. 
\ee     
(The full Kruskal manifold is illustrated in Fig.~4 below.)  Every point
represents a copy of three-dimensional Euclidean space ${\mathbb E}^3$
with coordinates $\bx$.  The metric is invariant under the full
$O(1,1)$ group including time reversal and space reflections.  It is
bounded on the left and right hand sides by the two hyperbolae
$r\equiv H= 0$, at which the metric is singular.  There are four
regions, denoted by I, II, III and IV. In regions II and IV, the
orbits of the boosts are spacelike. In region I they are
future-directed and timelike, and in region III they are past-directed
and timelike.  On the two Killing horizons $UV=0$, the orbits are
lightlike.  The two horizons cross on the Boyer axis at the origin
$U=0=V$.

\begin{figure}[ht]
\leavevmode\centering
\epsfxsize=5cm
\epsfbox{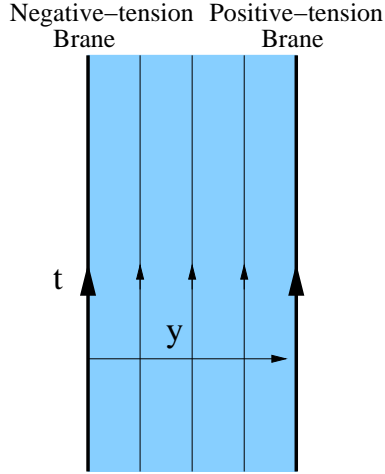}
\caption{\smallfont The time-independent Ho\v rava-Witten spacetime
in $(t,y)$ coordinates. Also shown are some typical orbits of the
timelike Killing vector field $\del/\del t$.}
\label{fig1a}
\end{figure}

\begin{figure}[ht]
\leavevmode\centering
\epsfxsize=5.69cm
\epsfbox{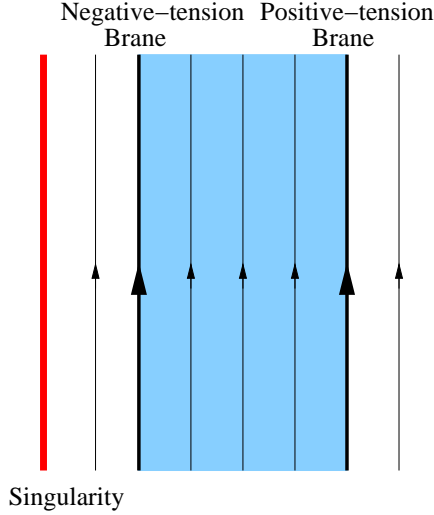}
\caption{\smallfont The full globally-static spacetime of the
time-independent Ho\v rava-Witten solution.  There is a single
singularity on the left, at $r=0$. Some representative orbits of
the timelike Killing vector field are indicated.
The Ho\v rava-Witten interval lies between two such orbits.}
\label{fig1b}
\end{figure}

   In the time-independent limit, that is $h=0$, the picture is rather
different.  The metric is globally static, and there are no
horizons. The spacetime is bounded by a single singularity at $r=0$.
The time independent Ho\v rava-Witten spacetime then occupies only a
portion of the full static manifold, between two values of $r$. In
Fig.~\ref{fig1a} we have sketched the Ho\v rava-Witten interval in
co-moving $(t,y)$ coordinates. It extends infinitely far in the
positive and negative time directions.  Also shown are some
representative orbits of the time translation Killing vector field. In
Fig.~\ref{fig1b} the Ho\v rava-Witten interval is shown lying between
two orbits of the static Killing vector field ${\partial \over
\partial t}$.

\begin{figure}[ht]
\leavevmode\centering
\epsfxsize=5.69cm
\epsfbox{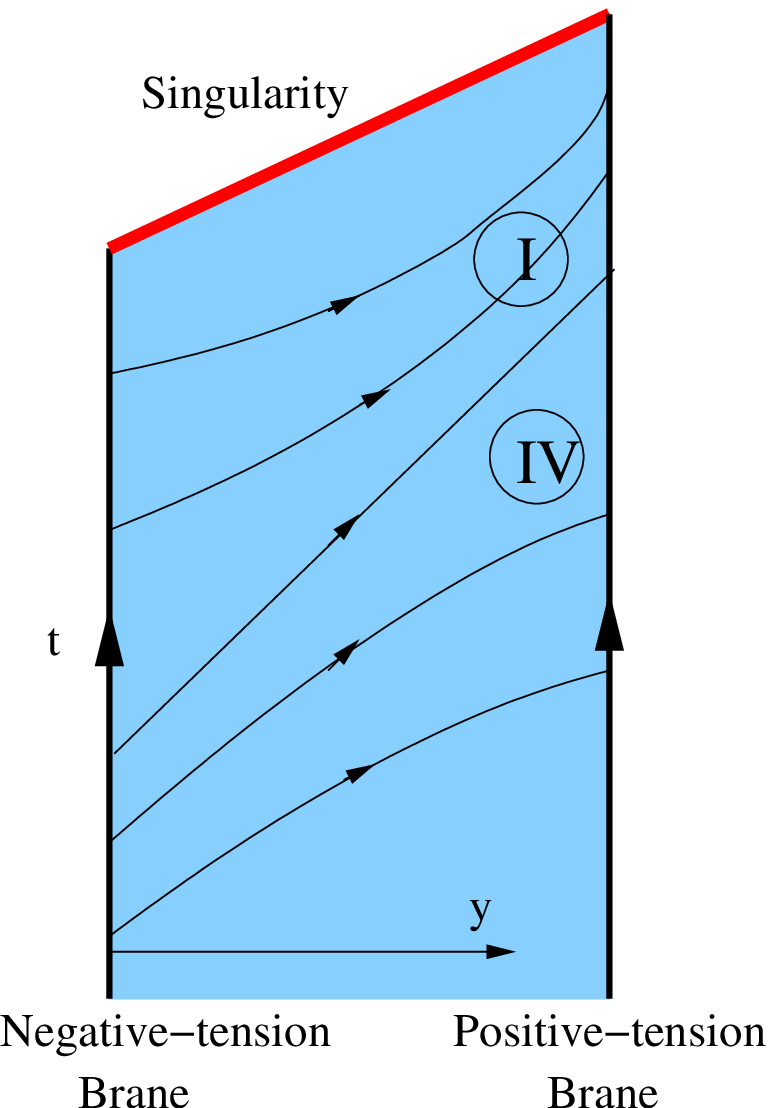}
\caption{\smallfont The time-dependent Ho\v rava-Witten spacetime
in co-moving $(t,y)$ 
coordinates.  The red line indicating the singularity occurs at $H=0$.
Also indicated are some of the orbits of the Killing vector 
$\del/\del t$. In region I the orbits are timelike and in region IV 
they are spacelike. These two regions are separated by a Killing horizon
on which the orbits are lightlike.}
\label{fig2a}
\end{figure}

   In Fig.~\ref{fig2a} we have shown the time-dependent Ho\v rava-Witten
interval in co-moving coordinates $(t,y)$. Also indicated are the
orbits of the timelike Killing field, which lie in the surfaces
$H={\rm constant}$.  In the past (\ie in region IV), it is spacelike.
In the future (\ie in region I), it is future-directed timelike.
These two regions are separated by a null surface -- the Killing
horizon.

\begin{figure}[ht]
\leavevmode\centering
\epsfxsize=8.85cm
\epsfbox{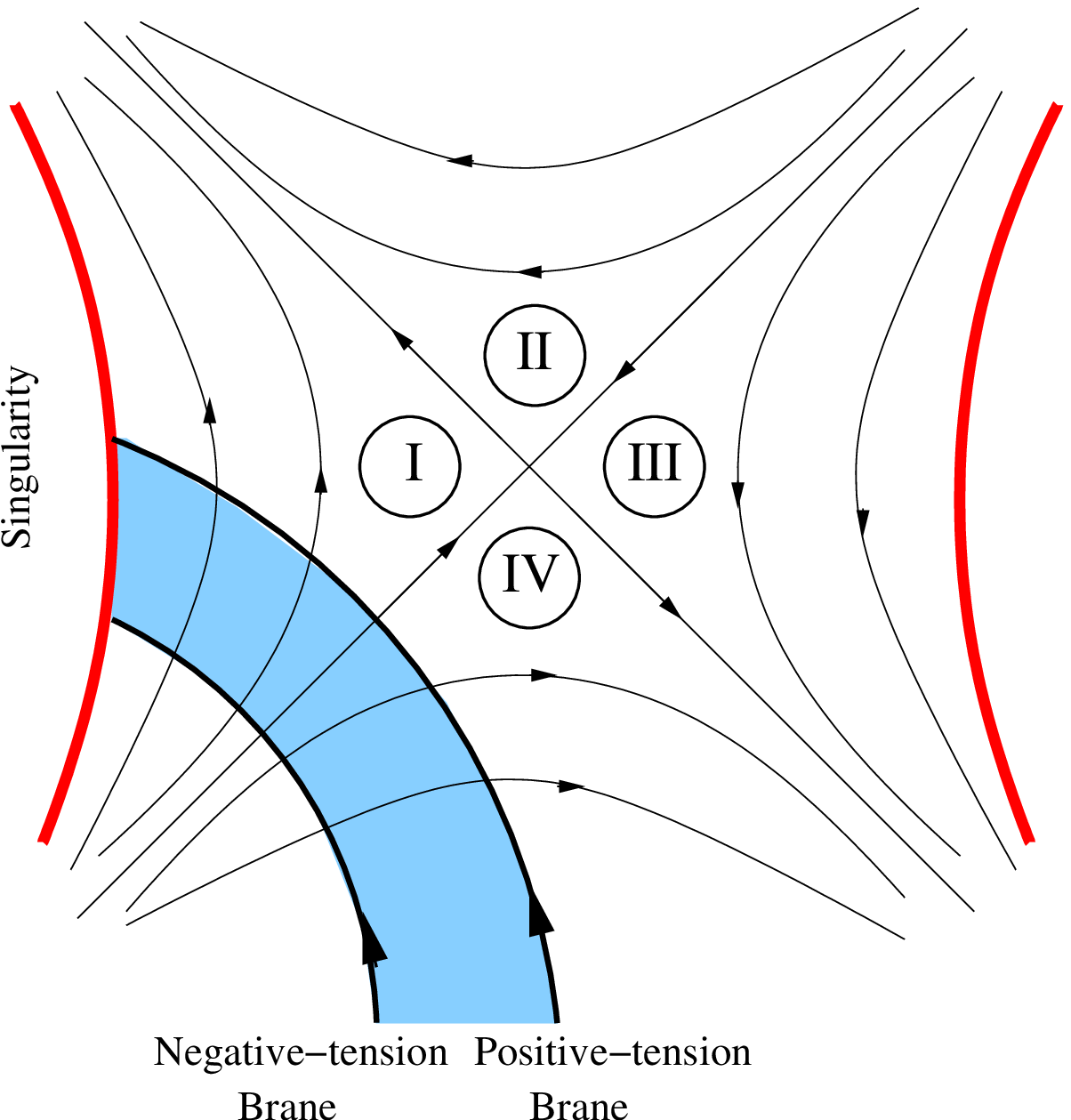}
\caption{\smallfont The full Kruskal manifold for the time-dependent
Ho\v rava-Witten
spacetime.  Some typical orbits of the Killing vector $\del/\del t$,
which are hyperbolae in the Kruskal coordinates $U$, $V$, are shown.
The orbits are future-directed timelike in region I, past-directed
timelike in region III, and spacelike in regions II and IV. Also
shown, shaded in blue, is the Ho\v rava-Witten manifold, bounded by the 
negative and positive tension branes, which move from region IV into region
I and then annihilate in the singularity.}
\label{fig2b}
\end{figure}

  In Fig.~\ref{fig2b} we have shown how the Ho\v rava-Witten interval is 
inserted into the full Kruskal manifold. It lies between two 3-branes,
which appear in the Kruskal diagram as two world lines, initially
starting in region IV, passing through the past horizon into region I,  
and coming to an end on the spacetime singularity at $r\equiv H=0$.     

   It is interesting that the global Kruskal spacetime picture
resembles the situation in de Sitter spacetime, except in that case
the origin $r=0$ is non-singular. The horizons then have more of the
character of cosmological horizons than black hole horizons, at least
for observers who remain in regions I or III. In particular, if we
reverse the direction of time, then the Ho\v rava-Witten interval
emerges from a spacetime singularity in the past and then inflates,
passing through a future Killing horizon. If we confine our attention
to the Ho\v rava-Witten interval however, this Killing horizon is not
a future event horizon for denizens of the negative tension brane or
indeed of the bulk.

\section{General Time-Dependent Domain-Wall Solutions}

    Here we obtain generalisations of our time-dependent solution
(\ref{tsol1}), to the case of domain-wall solutions in dilaton gravity
in arbitrary dimensions.

    Let us consider a general case of Einstein-scalar system with
an exponential potential:
\be
{\cal L} = \sqrt{-g} \Big(R - \ft12 (\del\phi)^2 -2 \Lambda\,
 e^{a\phi}\Big)\,,\label{genlag}
\ee
where the constant $a$ is parameterised by $a^2=\Delta +
\fft{2(D-1)}{D-2}$ \cite{pbrane}.
It is straightforward to verify that the system
admits the following solution
\bea
ds^2&=&H^{\ft8{(\Delta + 4)(D-2)}} \Big(-dt^2 + d\bx^2 + 
H^{\fft{2\Delta}{\Delta+4}}\, dy^2\Big)\,,\nn\\
\phi &=& -\fft{4a}{\Delta+4}\log H\,,\qquad
H=h\, t + k\, |y|\,,\label{newsol}
\eea
where $h$ is an arbitrary constant and $k$ is determined by
\be
k=\ft12 (\Delta+4)\sqrt{\fft{\Lambda}{\Delta}}\,.
\ee
It is interesting to note that for $\Delta>0$, which is typical in
{\it massive} supergravities, with $\Delta=4/N$ and $N$ taking integer
values from 1 to 8 depending on the dimensions\cite{pbrane,class},
$\Lambda$ is positive. Thus $k$ is real. On the other hand, if
$\Delta<0$, which is typical for {\it gauged} supergravities,
$\Lambda$ is negative.  This again implies that $k$ is real.  In
particular, for purely Einstein theory with a cosmological constant,
we have $\Delta=-2(D-1)/(D-2)$, the cosmological constant is negative,
and so $k$ is real.

    There are two ways of viewing the above solutions as
generalisations of previously-known solutions.  One is to show that
when $h=0$, the solutions reduce to standard domain-wall solutions,
supported by a purely exponential scalar potential.

   The other way of viewing the new solutions as generalisations of
old ones is to consider the limit $\Lambda=0$, under which the
solutions become Ricci-flat Kasner metrics with a $(D-2,1)$ spatial
splitting.

   Locally, where the modulus sign on $y$ is removed, 
it is possible to make the coordinate transformation
\bea
dt&=&d\td t - \fft{h\, r^{2\Delta/(\Delta+4)}}{k^2\,f} dr\,,
\qquad H=r\,,\nn\\
f&=&1 - \fft{h^2}{k^2}\, r^{\ft{2\Delta}{\Delta+4}}\,,
\label{coordtrans}
\eea
under which the solution (\ref{newsol}) becomes static, given by
\bea
ds^2 &=& r^{\ft{8}{(\Delta+4)(D-2)}}\Big(-f\, d\td t^2 +d\bx^2+
r^{\ft{2\Delta}{\Delta+4}}\fft{dr^2}{k^2\,f} \Big)\,,\nn\\
\phi &=&-\fft{4a}{\Delta+4}\log r\,,
\eea
This can be viewed as a static black brane.  As we discussed in
section \ref{timesec}, the mere fact that one can find a local
coordinate transformation that renders the metric static does not
imply that it is globally static, and in fact again, it is the modulus
sign on $y$ in the function $H$ that implies the solutions are
genuinely time dependent.

\subsection{AdS case}

   A special case of the above results is when $a=0$, in which case
the scalar-potential term in the Lagrangian (\ref{genlag}) becomes a
pure cosmological constant.  It corresponds to setting
$\Delta=-2(D-1)/(D-2)$.  The solution becomes
\be
ds^2 = H^{\fft4{D-3}}\, (-dt^2 + d\bx^2) + \fft{dy^2}{H^2}
\,,\label{adssol}
\ee
It follows from the curvature calculation (\ref{riemann}) that
if the solution is static, with $h=0$, the metric has only a
delta-function singularity at the location of the brane.  When
$h$ is non-vanishing, the metric has also a power-law singularity
at $H=0$.

    If one were to remove the modulus sign on $y$, the metric could
be locally transformed using (\ref{coordtrans}) into
the AdS black brane:
\be
ds^2=r^{-\ft4{D-3}} (-f\, dt^2 + d\bx^2) + \fft{dr^2}{k^2\, r^2\, f}
\,.
\ee
Again, it is the inclusion of the modulus sign that renders our
solution (\ref{adssol}) genuinely time-dependent.  Thus we see that
even anti-de Sitter spacetime itself is not immune to the Kasner-type
instabilities that we have exhibited in the general domain-wall
solutions, once the branes with their delta-function sources are
introduced.  This raises a question about the stability of the
Randall-Sundrum scenarios.

\section{Brane Sources}

    To obtain the full description of our solutions, with a source
brane action, it is useful first to dualise the cosmological constant
to a $D$-form field strength $F_{\sst D}= dA_{\sst{D-1}}$.  The full
action can be written as
\bea
I &=& \int d^Dx \sqrt{-g} \Big(R - \ft12 (\del \phi)^2 -
\ft1{2\, D!} e^{-a\phi} F_{\sst{(D)}}^2\Big) \nn\\
&& \qquad\qquad - T 
\int d^{D-1} \xi\sqrt{-\gamma} \Big(\gamma^{ij} \del_i x^M \del_j x^N
\td g_{MN} -(D-3) \Big)\nn\\
&&\qquad\qquad + \fft{(D-2)}{(D-1)!}\, \int d^{D-1}\xi\, 
   \varepsilon^{i_1\cdots i_{D-1}} \del_{i_1} x^{{M_1}}
\cdots \del_{i_{D-1}} x^{{M_{D-1}}} A_{M_1\cdots M_{D-1}}\,,
\label{fullact}
\eea
where $\td g_{MN}=e^{\ft{a}{D-1}\, \phi} g_{MN}$ is the metric in the
brane frame.\footnote{See \cite{dulukh} for a discussion of the
coupling of the dilaton in the brane action.}  This frame has the
defining property that the Einstein term and the $F_{\sst D}^2$ term
in the bulk Lagrangian have the same dilaton coupling, and the bulk
Lagrangian density takes the form
\be
{\cal L} = e^{-\ft12 a\phi\, (D-2)/(D-1)}\, \Big(\wtd R -
\ft1{2\, D!} F_{\sst{(D)}}^2 + \cdots \Big)\,.
\ee

   Varying the brane-action term in (\ref{fullact}) with respect to
the metric $g_{MN}$ gives a brane contribution to the energy momentum
tensor
\be
T_{\hbox{brane}}^{MN}=
T\, \int d^{D-1}\xi \sqrt{-\gamma}\, \gamma^{ij}\, e^{a\phi/(D-1)}\,
\del_i x^M\, 
\del_j x^N\, \fft{\delta^D(x-x(\xi))}{\sqrt{-g}}\,.
\ee
Making the static gauge choice $X^\mu=\xi^\mu$ ($0\le \mu \le D-2$)
implies that we have
\be
T_{\hbox{brane}}^{\mu\nu} = T\, e^{a\phi/(D-1)}\, 
\fft{\sqrt{-\gamma}}{\sqrt{-g}}\, \gamma^{\mu\nu}\delta(y)\,,
\ee
where $\gamma_{\mu\nu} = \td g_{\mu\nu}=  e^{a\phi/(D-1)}\,
g_{\mu\nu}$.  Substituting the solution (\ref{newsol}) into this
expression yields
\be
T^{\hbox{brane}}_{\mu\nu} = T\, H^{-(3\Delta+4)/(\Delta+4)}\,
\eta_{\mu\nu}\, \delta(y)\,.\label{tmunu}
\ee

   This singular brane source is precisely compatible with the Ricci
curvature singularities that we find for the metric in (\ref{newsol}),
resulting from the discontinuity in the gradient of $H$ at $y=0$.
Specifically, we find that there is a singular term in the Einstein
tensor, given by
\be
R_{\mu\nu}-\ft12 R\, g_{\mu\nu}  
= -\fft{8k}{\Delta+4}\, H^{-(3\Delta+4)/(\Delta+4)}\, 
        \eta_{\mu\nu}\, \delta(y) +\hbox{regular terms}\,.
\ee
Comparing with $T_{\mu\nu}$, we find that the powers of $H$ match 
precisely, and so the brane tension $T$ can then be read off as
\be
T=  -\fft{8k}{\Delta+4}\,.\label{branet}
\ee
This shows that the 3-brane at $y=0$ has negative tension.  If $y$ is
assigned period $2L$ and $y$ is identified with $-y$, the second
3-brane at $y=L$ in the resulting $S^1/{\mathbb Z}_2$ orbifold will
have positive tension.

   One can also check the dilaton equation of motion, which, including
the source term coming from the brane action in (\ref{fullact}), becomes 
\be
\square \phi + \fft{a}{2 D!}\, e^{-a\phi}\, F_{\sst{(D)}}^2 = 
          a T\, \fft{\sqrt{-\gamma}}{\sqrt{-g}}\, \delta(y)\,.
\ee
It is straightforward to verify that with the discontinuity in the
gradient of $\phi$ implied by (\ref{newsol}), the dilaton equation is
indeed satisfied, again with the brane tension $T$ given by
(\ref{branet}).

\section{Further Time-Dependent Solutions}

\subsection{Solutions with single-exponential potentials}

      We can obtain another type of time-dependent brane solution as
follows.  Consider the extremal static domain-wall solution for the
theory (\ref{genlag}).  By an appropriate choice of the $y$ coordinate
it can be written in the conformally-flat frame
\be
ds^2 = H^{\fft{2}{(\Delta +2)(D-2)}}
\Big(-dt^2 + dy^2 + d\bx^2\Big)\,,\label{confmet}
\ee
where $H=\td k\, y$, with $\td k=2(\Delta+2) k/(\Delta +4)$.
Now we can make a simple Lorentz boost, namely
$t\rightarrow c\, t + s\, y$ and $s\rightarrow s\, t + c\, y$, where
$c^2-s^2=1$, which implies that $H$ is now given by
\be
H= \td k s\, t + \td k c y\,.
\ee
Of course at this point there is no genuine time-dependence, since it
was merely obtained by a (globally-defined) coordinate transformation.

  If, however, we introduce a brane, by adding a modulus sign to $y$
and writing
\be
H= \td k s\, t + \td k c |y|\,,
\ee
then the time-dependence is no longer artificial, and the solution
describes a moving brane.  As usual, one could also extend to the
$S^1/{\mathbb Z}_2$ orbifold in the standard fashion.  An analysis of
the global structure of these solutions shows that they exhibit the
same essential behaviour as the previous examples, with the branes
moving towards each other and a power-law singularity developing that
eventually engulfs the entire spacetime.

\subsection{Scalar potential with an extremum}

So far we have considered supergravity domain-wall solutions
($(D-2)$-branes) supported by a single exponential potential, which
therefore has no extremum.  Here, we consider a more general scalar
potential that does have an extremum.  It is obtained from an $S^5$
reduction of type IIB supergravity in which the breathing mode is
retained, and the 5-form field strength has a non-trivial flux.  The
relevant scalar Lagrangian is given by \cite{instanton}
\be
{\cal L}_5 = \sqrt{-g} (R-\ft12 (\del\phi)^2 - V(\phi))\,,
\ee
with $V=8m^2 e^{8\alpha \phi} - R_5\, e^{16\alpha \phi/5}$ and
$\alpha= \sqrt{15}/12$.  Here $m$ measures the strength of the 5-form
flux and $R_5$ is the Ricci scalar of the internal $S^5$.  The scalar
potential supports a static domain-wall solution \cite{instanton}
\bea
ds_5^2 &=& (b_1 H^{2/7} + b_2 H^{5/7})^{-1/2}\, (-dt^2 + d\bx^2) +
 (b_1 H^{2/7} + b_2 H^{5/7})^{-2} dy^2\,,\nn\\
\phi &=& -\ft{\sqrt{15}}{7} \log H\,,\qquad
H=c + k|y|\,,\label{d5sol2}
\eea
where $b_1^2 = (28m/3k)^2$ and $b_2^2 = 196 R_5/45$.  The local
stability of the solution was recently analysed in \cite{lestsm},
where it was shown that subject to certain boundary conditions, the
configuration is stable despite the presence of the negative-tension
brane.

   However, as we have emphasised earlier, it is not clear that one is
entitled to impose the kind of energy-localising boundary conditions
that are needed in order to argue for stability, if one is considering
brane-world cosmological model.  Thus we may again look for a
time-dependent generalisation of the static solution, in order to
study the stability question from a viewpoint that is more in
accordance with the cosmological principle.  We find the following
time-dependent domain-wall solution:
\bea
ds_5^2 &=& (b_2 H^{3/7} + b_1)^{1/2} (ht (H^{3/7}-q) + q)^{1/3}
(-dt^2 + d\bx^2)\nn\\
&& + (b_2 H^{3/7} + b_1)^{-2}  (ht (H^{3/7}-q) + q)^{4/3}
H^{-8/7} dy^2\,,\nn\\
\phi &=& -\sqrt{\ft53} \log \Big(ht (H^{3/7} + q) -q \Big)\,,
\label{timedepsol}
\eea
where $q=-b_1/b_2$, and $h$ is an arbitrary constant.  The static
limit can be achieved by sending $t\rightarrow t + 1/h$ and then
sending $h\rightarrow 0$, whereupon the solution reduces to
(\ref{d5sol2}).

   Making the coordinate transformation $H^{3/7} = \fft{R_5^2}{400}
r^4 + q$, the solution (\ref{timedepsol}) can be written as
\bea
ds_5^2 &=& (W\,r^4)^{5/6} \Big(W^{-1/2} (-d\td t^2 + d\td\bx^2)
+ W^{1/2} dr^2\Big) \,,\nn\\
\phi&=&-\sqrt{\ft53} \log(h\td t\, r^4 + \td q)\,,\label{wsol}
\eea
where $W=h\td t + \td q/r^4$ and $\td q = 400q/R_5^2$, $\td x^\mu =
b_2^{1/4}\,x^\mu$.  The $r$ coordinate ranges over an interval $0 <
r_1\le r\le r_2$, where $r_1$ and $r_2$ are the values corresponding
to the brane locations at $y=0$ and $y=L$ respectively.  With $h$
taken to be negative, as usual, we again have the situation that the
solution is well-defined for sufficiently negative times $\td t$, with
the two 3-branes moving towards each other, but as $\td t$ increases
in the positive direction, a time is reached at which $W\le 0$ for all
$r_1\le r\le r_2$, at which point the annihilation of the universe is
complete.

   Lifting back to $D=10$, (\ref{wsol}) becomes 
\bea
ds_5^2&=& (h\td t + \fft{\td q}{r^4})^{-1/2} (-d\td t^2 + d\td \bx^2)
+ (h\td t + \fft{\td q}{r^4})^{1/2} (dr^2 + r^2 d\Omega_5^2)\,,\nn\\
F_5 &=& d^4\td x\wedge  dW^{-1} + {*d^4x \wedge dW^{-1}}\,.
\eea
This solution was obtained in \cite{glp}, describing a time-dependent
D3-brane.

       Analogous time-dependent domain-wall solutions can also be
found in $D=4$ and $D=7$, which can be obtained from $S^7$ and $S^4$
reductions of time-dependent M2-branes and M5-branes respectively.

\section{Conclusions}

     In this paper, we have constructed time-dependent solutions of
dilaton gravity with an exponential potential, which can be viewed as
generalisations of the static domain-wall solutions of supergravities
in various dimensions.  Included in these solutions are time-dependent
generalisations of the five-dimensional heterotic 3-brane that was
proposed in \cite{Ovrutetal} as a model for our universe, and of the
AdS 3-brane of the Randall-Sundrum scenarios.  The case of principle
interest is where the fifth dimension is a line segment, with a
positive-tension brane at one end and a negative-tension brane at the
other.  The time-dependent solution starts out in the distant past in
a non-singular regime in which the metric approximates a Kasner model
far from the Kasner singularity.  The solution evolves to a
singularity in which the entire spacetime is annihilated.  The
existence of this time-dependent solution can be taken as an
indication of an inherent classical instability in brane-world models
where there are positive-tension and negative-tension branes present.
Analogous conclusions can be drawn for brane models of this type in
other dimensions, and also in certain more general cases where there
is a scalar potential with an extremum.  Some support for the
idea that  Kasner singularities arise in the  general case
is given by the numerical work reported in \cite{mafefrkope}.
Further support comes from $(2+1)$-dimensional models \cite{martro}.
The thermodynamics of negative tension branes has been discussed
in \cite{marros}.

     The singular behaviour of solutions may be ascribed to a failure
to stabilise some of the modulus fields of the compactification.  This
is an old problem, and we have little that is new to say about it.  By
introducing a potential (as for example in \cite{KKLT,becukr,Beckers}) 
the situation should improve. However in the cases we
have considered the modulus problem is compounded by the presence of
negative tension branes. Such negative tension branes are almost
inevitable in any warped compactification \cite{Linde}.  It is by no
means obvious that they can be stabilised by additional
potentials. Moreover, ideally, one would want not so much a stable
static solution, but rather a stable Friedman-Lemaitre expanding
solution, and, ideally, it should behave like an ``attractor,'' as was
attempted in the early days of Kaluza-Klein cosmology
\cite{Chapline,Maeda}.  In this respect, an unstable static solution
might be thought of as an advantage since otherwise the universe could
be trapped in limbo in a stable static universe.  Unfortunately at
present no higher-dimensional field equations using the potentials
found in \cite{KKLT} and \cite{becukr,Beckers} appear to be available to test
this idea.

\section*{Acknowledgements}

    We should like to thank Neil Turok for helpful discussions, 
and Harvey Reall for drawing our attention to \cite{ChamblinReall}.

\section*{APPENDIX}

\appendix

\section{Curvature of the Domain-Wall Metrics}

   All the domain-wall metrics that we have been considering in this
paper have the general form
\be
ds^2 = H^{2\alpha}\, (-dt^2 + d\bx^2 ) + H^{2\beta}\, dy^2\,,
\ee
where $\alpha$ and $\beta$ are constants.  We shall make the obvious
choice of orthonormal frame, defining
\be
e^0 = H^\alpha\, dt\,,\qquad e^i= H^\alpha\, dx^i\,,\qquad
   e^y = H^{\beta}\, dy\,.
\ee
The orthonormal frame  components of the Riemann tensor are then given
by
\bea
R_{0i0j} &=& \alpha\, H^{-2\alpha}\, \Big( \fft{\dot H^2}{H^2} -
    \fft{\ddot H}{H}\Big)\, \delta_{ij} + \alpha^2\, H^{-2\beta-2}\,
        {H'}^2\, \delta_{ij}\,,\nn\\
R_{0ijy} &=& \alpha\, H^{-\alpha-\beta}\, \Big(\fft{\dot H'}{H} -
    (1+\beta)\, \fft{\dot H\, H'}{H^2}\Big)\, \delta_{ij}\,,\nn\\
R_{0y0y} &=& \alpha\, H^{-2\beta}\, \Big( \fft{H''}{H}+
(\alpha-\beta-1)\,
    \fft{{H'}^2}{H^2}\Big) -\beta\, H^{-2\alpha}\, \Big( \fft{\ddot
      H}{H} + (\beta-\alpha-1)\, \fft{\dot H^2}{H^2}\Big)\,,\nn\\
R_{ijk\ell} &=& \alpha^2\, \Big( H^{-2\alpha}\, \fft{{\dot H}^2}{H^2}
      - H^{-2\beta}\, \fft{{H'}^2}{H^2}\Big)\, (\delta_{ik}\,
        \delta_{j\ell} - \delta_{i\ell}\, \delta_{jk})\,,\nn\\
R_{iyjy} &=& \alpha\, \beta\, H^{-2\alpha-2}\, {\dot H}^2\,
\delta_{ij}  -\alpha\, H^{-2\beta}\, \Big(\fft{H''}{H} +
       (\alpha-\beta -1)\, \fft{{H'}^2}{H^2}\Big)\, \delta_{ij}\,.
\label{riemann}
\eea


\begin{thebibliography}{99} 

\bibitem{H-W}P. Ho\v rava and E. Witten,
{\it Heterotic and type I string dynamics from eleven dimensions,}
Nucl. Phys. {\bf B460}, 506 (1996), hep-th/9510209.

\bm{lalupo} I.V. Lavrinenko, H. L\"u and C.N. Pope,
{\it From topology to generalised dimensional reduction}, 
Nucl. Phys. {\bf B492}, 278 (1997), hep-th/9611134.

\bibitem{Ovrutetal} A. Lukas, B.A. Ovrut, K.S. Stelle and D. Waldram,
{\it The universe as a domain wall}, 
Phys. Rev. {\bf D59}, 086001 (1999), hep-th/9803235.


\bibitem{Ek} J. Khoury, B.A. Ovrut, P.J. Steinhardt and N. Turok, {\it
The ekpyrotic universe: Colliding branes and the origin of the hot big
bang,} Phys. Rev. {\bf D64}, 123522 (2001), hep-th/0103239.

\bibitem{cyclic}P.J. Steinhardt and N. Turok,
{\it Cosmic evolution in a cyclic universe,} Phys. Rev. {\bf D65}, 
126003 (2002), hep-th/0111098.

\bibitem{Ishibashi} 
G.W. Gibbons and A. Ishibashi,
{\it Topology and signature changes in braneworlds},
Class. Quant. Grav.  {\bf 21}, 2919 (2004), hep-th/0402024.

\bibitem{glp} G.W. Gibbons, H. L\"u and C.N. Pope,
{\it Brane worlds in collision,} hep-th/0501117.

\bm{kastra} D. Kastor and J.H. Traschen,
{\it Cosmological multi-black-hole solutions,}
Phys. Rev. {\bf D47}, 5370 (1993), hep-th/9212035.

\bm{makshi} T. Maki and K. Shiraishi, {\it Multi-black hole solutions
in cosmological Einstein-Maxwell dilaton theory,} Class. Quant. Grav.
{\bf 10}, 2171 (1993).


\bibitem{KKLT} S. Kachru, R. Kallosh, A.D. Linde and S.P. Trivedi,
{\it de Sitter vacua in string theory,}
Phys.\ Rev.\ {\bf D68}, 046005 (2003), hep-th/0301240.

\bm{becukr} M. Becker, G. Curio and A. Krause,
{\it de Sitter vacua from heterotic M-theory},
Nucl. Phys. {\bf B693}, 223 (2004), hep-th/0403027.

\bibitem{Beckers} K. Becker, M. Becker and A. Krause,
{\it M-theory inflation from multi M5-brane dynamics},  hep-th/0501130.

\bibitem{Carter} B. Carter, {\it Black hole equilibrium states}, in
{\sl Black Holes (Les Houches Lectures)}, eds. B.S. DeWitt and
C. DeWitt (Gordon and Breach, New York, 1972).

\bibitem{Gibbons} G.W. Gibbons, 
{\it Global structure of supergravity domain wall space-times},
Nucl. Phys.  {\bf B  394}, 3 (1993).

\bibitem{Whewell} W. Whewell, {\sl Review of Volume II of Lyell's
Principles of Geology}, in {\it Quarterly Review} {\bf 48}, 126 (1832).

\bibitem{ChamblinReall}
H.A. Chamblin and H.S. Reall,
{\it Dynamic dilatonic domain walls},
Nucl Phys. {\bf B 562}, 133 (1999), hep-th/9903225.

\bibitem{pbrane} H. L\"u and C.N. Pope,
{\it p-brane solitons in maximal supergravities,}
Nucl.\ Phys.\ {\bf B465}, 127 (1996), hep-th/9512012.

\bibitem{class} H. L\"u, C.N. Pope, T.A. Tran and K.W. Xu,
{\it Classification of p-branes, NUTs, waves and intersections,}
Nucl.\ Phys.\ {\bf B511}, 98 (1998), hep-th/9708055.

\bibitem{instanton} M.S. Bremer, M.J. Duff, H. L\"u, C.N. Pope
and K.S. Stelle, {\it Instanton cosmology and domain walls from
M-theory and string theory,} Nucl. Phys. {\bf B543}, 321 (1999),
hep-th/9807051.

\bm{dulukh}M.J. Duff, R.R. Khuri and J.X. Lu,
{\it String solitons},
Phys. Rept. {\bf 259}, 213 (1995), hep-th/9412184.

\bibitem{lestsm} J.L. Lehners, K.S. Stelle and P. Smyth,
{\it Stability of Ho\v rava-Witten spacetimes,} hep-th/0501212.

\bm{mafefrkope} J. Martin, G.N. Felder, A.V. Frolov, L. Kofman and M. Peloso,
{\it BRANECODE: A program for simulations of braneworld dynamics}, 
hep-ph/0404141.

\bm{martro} D. Marolf and M. Trodden,
{\it Black holes and instabilities of negative tension branes},
Phys. Rev. {\bf D64}, 065019 (2001), hep-th/0102135.

\bm{marros} D. Marolf and S.F. Ross,
{\it Stringy negative-tension branes and the second law of thermodynamics},
JHEP {\bf 0204}, 008 (2002), hep-th/0202091.

\bibitem{Linde}
G.W. Gibbons, R. Kallosh and A.D. Linde,
{\it Brane world sum rules} 
JHEP {\bf 0101}, 022 (2001), hep-th/0011225

\bibitem{Chapline} 
G.F. Chapline and G.W. Gibbons,
{\it Unification of elementary particle physics and cosmology in
ten dimensions}, Phys. Lett. {\bf B 135}, 43 (1984).

\bibitem{Maeda} K. Maeda,
{\it Stability and attractor in a higher dimensional cosmology: II},
 Class Quant Grav  {\bf 3 }, 651  (1986).


\end{thebibliography}
\end{document}